\documentclass[twocolumn,showpacs,preprintnumbers,amsmath,amssymb]{revtex4}

\usepackage{graphicx}
\usepackage{dcolumn}
\usepackage{bm}

\input epsf

\begin{document}


\title{Nernst effect in the vortex-liquid regime of a type-II superconductor}

\author{Subroto Mukerjee}
\email{mukerjee@princeton.edu}
\author{David A. Huse}
\email{huse@princeton.edu}
\affiliation{Department of Physics, Princeton University, Princeton, NJ 08544}

\date{\today}

\begin{abstract}
We measure the transverse thermoelectric coefficient $\alpha_{xy}$
in simulations of type-II superconductors in the vortex liquid
regime, using the time-dependent Ginzburg-Landau (TDGL) equation
with thermal noise. Our results are in reasonably good
quantitative agreement with experimental data on cuprate samples,
suggesting that this simple model 
contains much of the physics behind the large Nernst effect due to
superconducting fluctuations observed in these materials.
\end{abstract}

\pacs{74.25.Fy, 72.15.Jf}
\maketitle

\section{Introduction}

The Nernst effect in the cuprate superconductors has recently
become a focus of attention both experimentally
\cite{palstra,ri,ong1,ong2,ong3} and theoretically
\cite{ud,iddo,iddo2,tl}.  The Nernst effect is the electric field
induced when the sample has a temperature gradient,
${\bm\nabla}T$, perpendicular to the magnetic field, ${\bf H}$;
this electric field is perpendicular to both ${\bm\nabla}T$ and
${\bf H}$.  For some cuprate superconductors the Nernst effect due
to superconducting fluctuations is detectable at temperatures far
above the transition temperature, $T_c$ \cite{ong1, ong2}.
Ussishkin {\it et al.} \cite{iddo} calculated the low-field Nernst
effect for $T>T_c$ due to superconducting fluctuations, obtaining
results in reasonable agreement (in absolute units) with
experimental data. They used the linearized time-dependent
Ginzburg-Landau (TDGL) equation, which is identical to the
Aslamazov-Larkin \cite{aslamazov} approximation for the
microscopics.  Linearized (Gaussian) TDGL is applicable far enough
above the mean-field transition temperature $T_{c}^{MF}$, where
the order parameter fluctuations are small enough to neglect the
nonlinear terms in the full Ginzburg-Landau free energy.  To
estimate the Nernst effect closer to $T_c$ they also treated the
nonlinearities in TDGL in the Hartree approximation \cite{iddo}.

Prominent in the phase diagrams of the cuprate superconductors in
a magnetic field is the {\it vortex liquid regime}, where
short-range superconducting correlations are strong (because
$T<T_c$), but thermal phase and vortex fluctuations disrupt the
superconducting coherence on longer scales. The experimentally
observed Nernst effect remains large in this regime
\cite{ri,ong1,ong2,ong3}, which
is not accessible to analytic calculations due to its strong
correlations and fluctuations.  However, the TDGL equation with
its full nonlinearity and driven by thermal fluctuations can be
numerically simulated in this vortex liquid regime, and its Nernst
effect measured and compared to experiments, as we demonstrate
below.

When the TDGL approximation is used to model properties of cuprate
superconductors, for which there is not a well-established more
microscopic theory, or more generally when it is used well away
from the zero-field critical temperature, as we do here, it
constitutes a standard, widely-used, but essentially
phenomenological approximation. It has the virtue that most of the
parameters that enter can be determined from experiments, as we
discuss below. For BCS superconductors, the TDGL approximation can
be systematically derived from the microscopics in some limited
regimes \cite{tink,larkin}, but here we are instead using it more
broadly, in regimes where its justification is ``only''
phenomenological.

The Nernst effect in the vortex liquid can be described in terms
of the vortices as the phase-slip voltage due to vortices being
transported down the temperature gradient as heat carriers.
However, when one calculates the Nernst effect at higher
temperatures in linearized TDGL \cite{iddo}, what is happening is
that superconducting order parameter fluctuations are transported
along the temperature gradient and their phase patterns are
twisted by this motion across the magnetic field, inducing a
Nernst voltage; vortices play no role in the calculation.  In a
TDGL treatment of the vortex liquid regime, there is no clean
distinction between order parameter fluctuations and vortex
motion, so both of these two different-sounding descriptions are
in some sense correct.

In this paper, we present the method for and results of
simulations of thermoelectric transport in type-II superconductors
in the vortex liquid regime.  We simulate the TDGL equation with
thermal noise.  We work in the strongly type-II limit, $\kappa\gg
1$, where the magnetic field in the sample is assumed to be
uniform and not fluctuating.  Mostly we work in two dimensions,
but we also examine the crossover from two-dimensional to
three-dimensional behavior. In our study of interlayer couplings
we see some indication that a substantial part of the entropy
carried by the vortices in the vortex liquid may be the
configurational entropy of their positions.

Our simulations involve a dimensionless tunable parameter ($\eta$)
which sets the strength of the thermal fluctuations in the sample
and hence can range from nearly mean-field behavior with very weak
fluctuations to very strong fluctuations with strongly suppressed
superconductivity. We show that for an intermediate value of this
parameter our results are in reasonably good qualitative and
quantitative agreement with available experimental Nernst-effect
data on overdoped La$_{2-x}$Sr$_{x}$CuO$_{4}$ (LSCO) \cite{ong3}.
In this comparison all the other parameters in the simulations are
set by experimentally measured quantities.  We also discuss in
brief the results of our simulations to model experiments on
Bi$_{2}$Sr$_{2}$CaCu$_{2}$O$_{8+x}$ (BSCCO) \cite{ri}.

\section{Some thermo-electro-magnetic transport theory}

We briefly review some of the general theory of thermal and
electrical transport coefficients in linear response in the
presence of a magnetic field. Consider a sample at temperature $T$
subjected to small gradients in the potential ${\bf \nabla} \phi$
and temperature ${\bf \nabla} T$.
One can then very generally write down the transport current
densities in the system to linear order in ${\bf \nabla} \phi$ and
${\bf \nabla} T$ as
\begin{equation}
{\bf J}^{e}_{\rm tr} = -\hat{\sigma} {\bf \nabla} \phi -
\hat{\alpha}{\bf \nabla} T ~,
\end{equation}
\begin{equation}
{\bf J}^{Q}_{\rm tr} = \hat{\tilde{\alpha}}{\bf \nabla} \phi -
\hat{\kappa}{\bf \nabla} T ~,
\end{equation}
where ${\bf J}^{e}_{\rm tr}$ is the charge transport current
density and ${\bf J}^{Q}_{\rm tr}$ is the heat transport current
density. $\hat{\sigma}$, $\hat{\alpha}$, $\hat{\tilde{\alpha}}$
and $\hat{\kappa}$ are the electrical conductivity,
thermoelectric, electrothermal and thermal conductivity tensors.
The Onsager relations for the transport coefficients tell us that
$\hat{\tilde{\alpha}}=T\hat{\alpha}$. The Nernst coefficent
($\nu$) can be defined in a configuration with a magnetic field
${\bf B \| \hat{z}}$ and ${\bf \nabla T \| \hat{x}}$ along with
the condition ${\bf J}^{e}_{\rm tr} = 0$ as
\begin{equation}
\nu = -\frac{E_{y}}{B \nabla T} =
\frac{1}{B}\frac{\alpha_{xy}\sigma_{xx}-\alpha_{xx}\sigma_{xy}}{\sigma_{xx}^{2}+\sigma_{xy}^{2}}
~.
\end{equation}
If the system shows no Hall effect (like the one we have
simulated), then $\sigma_{xy}=0$ and
\begin{equation}
\nu = \frac{\alpha_{xy}}{B\sigma_{xx}} ~.
\end{equation}

One way to obtain transport coefficients from a simulation is via
the Kubo relations, thus measuring current-current correlations in
the equilibrium dynamics.  Alternatively, one can apply an
electric field and/or a temperature gradient and measure the
resulting currents. We have used both methods, finding that our
implementation of the latter method gives higher signal-to-noise
ratio per unit computer time in the regime we have been studying.
The currents flowing in the sample have magnetization parts in
addition to the transport parts considered above. One could write
down linear response relations for the total current densities
analogous to the ones for the transport current densities but the
coefficients appearing in those would not obey the Onsager
relations. The total currents in the system can in general be
computed in a straightforward manner but one has to extract the
transport parts from these to calculate the transport
coefficients. This involves identifying and subtracting out the
magnetization currents. The procedure we use to that end in this
work is outlined below and is based on the arguments of Cooper
{\it et al.} ~\cite{halperin}.

Let ${\bf J}^{e}_{\rm tot}({\bf r})$ and ${\bf J}^{Q}_{\rm
tot}({\bf r})$ be the total charge and heat current densities
(transport + magnetization) at any point ${\bf r}$ in the sample.
If $\phi({\bf r})$ is the electric potential at ${\bf r}$, there
exists a total energy current ${\bf J}^{E}_{\rm tot}({\bf r})$
such that
\begin{equation}
{\bf J}^{Q}_{\rm tot}({\bf r}) = {\bf J}^{E}_{\rm tot}({\bf r}) -
\phi({\bf r}){\bf J}^{e}_{\rm tot}({\bf r}) ~.
\end{equation}
A similar relation holds between the transport parts of these
current densities,
\begin{equation}
{\bf J}^{Q}_{\rm tr}({\bf r}) = {\bf J}^{E}_{\rm tr}({\bf r}) -
\phi({\bf r}){\bf J}^{e}_{\rm tr}({\bf r}) ~.
\end{equation}
The average electric current density ${\bf\bar J}^{e}_{\rm tr}$
and heat current density ${\bf\bar J}^{Q}_{\rm tr}$ are given by
averaging ${\bf J}^{e}_{\rm tr}({\bf r})$ and ${\bf J}^{Q}_{\rm
tr}({\bf r})$ over the whole sample. For a two dimensional sample
\begin{equation}
{\bf\bar J}^{e}_{\rm tr} = \frac{1}{S}\int_{S}{\bf J}^{e}_{\rm
tr}({\bf r})dS
\end{equation}
and
\begin{equation}
{\bf\bar J}^{Q}_{\rm tr} = \frac{1}{S}\int_{S}{\bf J}^{Q}_{\rm
tr}({\bf r})dS ~,
\end{equation}
where $S$ is the area of the sample. Similar relations also hold
for three dimensional samples. From Eqns. (6) and (8), we obtain
\begin{equation}
{\bf\bar J}^{Q}_{\rm tr} = \frac{1}{S}\left(\int_{S}{\bf
J}^{E}_{\rm tr}({\bf r})dS - \int_{S}\phi({\bf r}){\bf J}^{e}_{\rm
tr}({\bf r})dS\right) ~.
\end{equation}
We know that
\begin{eqnarray}
{\bf J}^{e}_{\rm tot}({\bf r}) & = & {\bf J}^{e}_{\rm tr}({\bf r}) + {\bf J}^{e}_{\rm mag}({\bf r}) \\
{\bf J}^{E}_{\rm tot}({\bf r}) & = & {\bf J}^{E}_{\rm tr}({\bf r})
+ {\bf J}^{E}_{\rm mag}({\bf r}) ~, \nonumber
\end{eqnarray}
where ${\bf J}^{e}_{\rm mag}({\bf r})$ and ${\bf J}^{E}_{\rm
mag}({\bf r})$ are the charge and energy magnetization current
densities. It can be shown on general grounds ~\cite{halperin}
that there exist charge and energy magnetization densities ${\bf
M}^{e}({\bf r})$ and ${\bf M}^{E}({\bf r})$ such that
\begin{eqnarray}
{\bf J}^{e}_{\rm mag}({\bf r}) & = & {\bf \nabla} \times {\bf M}^{e}({\bf r})\\
{\bf J}^{E}_{\rm mag}({\bf r}) & = & {\bf \nabla} \times {\bf
M}^{E}({\bf r}) ~.\nonumber
\end{eqnarray}

The charge magnetization of the system is nothing but the
conventional magnetic moment density ${\bf M}({\bf r})$. If the
material surrounding the sample is assumed to be non-magnetic,
both ${\bf M}^{e}({\bf r})$ and ${\bf M}^{E}({\bf r})$ vanish
outside the sample. The magnetization currents are mostly at the
boundaries because that is where the change in magnetization is
the largest. Thus we see that
\begin{eqnarray}
\int_{S}{\bf J}^{e}_{\rm tr}({\bf r})dS & = & \int_{S}{\bf J}^{e}_{\rm tot}({\bf r})dS \\
\int_{S}{\bf J}^{E}_{\rm tr}({\bf r})dS & = & \int_{S}{\bf
J}^{E}_{\rm tot}({\bf r})dS ~. \nonumber
\end{eqnarray}
From Eqns (9) and (12), we obtain
\begin{equation}
{\bf\bar J}^{Q}_{\rm tr} = \frac{1}{S}\left(\int_{S}{\bf
J}^{E}_{\rm tot}({\bf r})dS - \int_{S}\phi({\bf r}){\bf
J}^{e}_{\rm tr}({\bf r})dS\right) ~.
\end{equation}
It should be noted that there is no ``heat magnetization density"
${\bf M}^{Q}({\bf r})$ analogous to ${\bf M}^{e}({\bf r})$ and
${\bf M}^{E}({\bf r})$ for which the heat magnetization current
${\bf J}^{Q}_{\rm mag}({\bf r}) = {\bf \nabla} \times {\bf
M}^{Q}({\bf r})$. A consequence of this is that unlike ${\bf
J}^{e}$ and ${\bf J}^{E}$,
$$\int_{S}{\bf J}^{Q}_{\rm tr}({\bf r})dS \neq \int_{S}{\bf J}^{Q}_{\rm tot}({\bf r})dS ~.$$
In fact, it can be shown that
\begin{equation}
{\bf J}^{Q}_{\rm mag}({\bf r}) = {\bf \nabla}\phi \times {\bf
M}({\bf r}) ~.
\end{equation}

\section{Simulations}

We measure the transport coefficient $\tilde{\alpha}_{xy}$ by
turning on an electric field $-{\bf \nabla} \phi {\bf \| \hat{y}}$
and a magnetic field ${\bf B \| \hat{z}}$ and measuring $\bar
J^{Q(x)}_{\rm tr}$, the $x$ component of the average heat
transport current density at constant temperature.  This is done
in a system with mixed boundary conditions: it has free surfaces
at the boundaries normal to $\hat{y}$, while it has periodic
boundary conditions along the $x$-direction (and along $\hat{z}$
when we study 3D). If, as in our model, the system does not have a
Hall response, under these conditions the $x$ component of the
electric transport current vanishes at all points, $J^{e(x)}_{\rm
tr} ({\bf r}) = 0$. Thus from Eqns. (2) and (13), we obtain
\begin{equation}
\tilde{\alpha}_{xy} = -\frac{\bar J^{Q(x)}_{\rm tr}}{\nabla_{y}
\phi} = -\frac{1}{S}\frac{\left(\int_{S}J^{E(x)}_{\rm tot}({\bf
r})dS\right)}{\nabla_{y} \phi} ~.
\end{equation}
Similarly, $\alpha_{xy}$ can be measured by introducing a
temperature gradient ${\bf \nabla} T {\bf \| \hat{y}}$ and
magnetic field ${\bf B \| \hat{z}}$ with no external electric
potential and measuring $\bar J^{e(x)}_{\rm tr}$. From Eqns. (1)
and (12),
\begin{equation}
\alpha_{xy} = -\frac{\bar J^{e(x)}_{\rm tr}}{\nabla_{y} T}  =
-\frac{1}{S}\frac{\left(\int_{S}J^{e(x)}_{\rm tot}({\bf
r})dS\right)}{\nabla T} ~.
\end{equation}
The two coefficients $\tilde{\alpha}_{xy}$ are $\alpha_{xy}$ can
thus be obtained by measuring only the total currents in the
system. They are related to each other by the Onsager relation and
to the Nernst coefficient by Eqn. (4).


The TDGL equation that we have simulated is
\begin{eqnarray}
\tau \left( \partial_{t}+i \frac{e^{*}}{\hbar} \phi \right)
\Psi & = & \frac{\hbar^{2}}{2m^{*}} \left( \bm{\nabla}-i \frac{e^{*}}{\hbar}
{\bf A} \right)^{2} \Psi \\
& & - a_{0}(T-T^{MF}_{c})\Psi - b |\Psi|^{2}\Psi + \zeta({\bf r},t)~. \nonumber
\end{eqnarray}
We take $\tau$ to be real so there is neither Hall effect nor
Seebeck effect \cite{ud}. This approximation of leaving out the
Hall effect should be reasonable as long as the Hall angle is
small, as it generally is in the vortex liquid regime of the
cuprate superconductors.  We work in the type-II limit where the
magnetic field is assumed to be uniform and not fluctuating. The
noise correlator is
\begin{equation}
<\zeta^{*}({\bf r'},t')\zeta({\bf r},t)> = 2\tau k_{B}T
\delta({\bf r}-{\bf r'})\delta(t-t') ~.
\end{equation}
The current operators we need are \cite{ud}
\begin{eqnarray}
{\bf J}^{e}_{\rm tot} & = & -i\frac{e^{*}\hbar}{2m^{*}} \langle \Psi^{*}\left(\bm{\nabla} - \frac{i e^{*}}{\hbar}{\bf A} \right) \Psi \rangle + c.c. \\
{\bf J}^{E}_{\rm tot} & = & -\frac{\hbar^{2}}{2m^{*}} \langle \frac{\partial\Psi^{*}}{\partial t} \left(\bm{\nabla} - \frac{i e^{*}}{\hbar}{\bf A} \right) \Psi \rangle + c.c. \nonumber
\end{eqnarray}

The TDGL equation is the simplest dynamical stochastic equation
one can obtain from the Ginzburg-Landau free energy functional. As
a consequence of its simple relaxational dynamics, it does not
explicitly conserve either total charge or total energy.  It is
implicitly in contact with and exchanging energy and charge with
local reservoirs.  In reality these reservoirs are presumably the
phonons, quasiparticles and other excitations not included in the
TDGL equation.  Nevertheless, the charge and energy currents
carried by the superconducting order parameter $\Psi$ and its
fluctuations are as given above and can be measured and their
transport properties determined within this model. Charge and energy
conservation can explicitly be taken into account in microscopic
derivations of the TDGL theory, where only the contribution due to
superconducting fluctuations is retained. It can however be shown that other
effects (normal state contributions etc.) do not contribute as significantly
to the Nernst effect as superconducting fluctuations \cite{iddo2}.

An important feature of TDGL for the present work is that the
parameter $\tau$, which sets the time scale, does not enter in the
values of the transverse thermoelectric coefficients that we are
studying, $\alpha_{xy}$ and $\tilde \alpha_{xy}$. This allows
quantitative comparison to experiments to be done without
estimating $\tau$, which is fortunate, because the value of $\tau$
is quite uncertain.

This TDGL equation with noise has a dimensionless parameter that gives the strength of the thermal fluctuations.  To remove all the dimensional quantities, we use $\xi_{0}$, the zero temperature coherence length, as our unit of length;  $T_{c}^{MF}$, $k_BT_c^{MF}$, $\frac{\hbar}{e^*}$ and  $\tau/(a_{0}T_{c}^{MF})$ as our units of temperature, energy, magnetic flux and time, respectively; and $\Psi_0$, the zero-temperature order parameter magnitude, as our unit for the order parameter.  The resulting TDGL equation is then
\begin{equation}
\left( \partial_{t}+i \phi \right) \Psi = \left(\bm{\nabla}-i{\bf A} \right)^{2}\Psi -(T-1)\Psi -|\Psi|^{2}\Psi + \zeta({\bf r},t)~,
\end{equation}
with noise correlator
\begin{equation}
<\zeta^{*}({\bf r'},t')\zeta({\bf r},t)> = 2\eta T \delta({\bf r}-{\bf r'})\delta(t-t')~,
\end{equation}
and
\begin{equation}
\eta=\frac{bk_BT_c^{MF}}{\xi_0^d(a_0T_c^{MF})^2}
\end{equation}
is the fluctuation parameter, where $d$ is the number of spatial
dimensions.  When $\eta \ll 1$, the actual critical temperature
$T_c$ is near the mean-field critical temperature $T_c^{MF}$,
while if $\eta \gg 1$ then $T_c \ll T_c^{MF}$. The model has four
parameters that remain after scaling to the units specified above:
the temperature, the magnetic field, the cutoff, and the strength
$\eta$ of the noise.

We initially simulate a two-dimensional system, thus ignoring
interlayer coupling.  We discretize space and time.  The spatial
grid spacing is taken to be $\xi_{0}$.  This rather coarse spatial
grid is used to minimize the computer time needed to simulate
samples large compared to this microscopic length. There are
noticeable quantitative effects of using such a coarse grid. For
example, in our units $H_{c2}(T=0) \cong 1.18$ with this grid,
while it is 1.0 for the continuum model.  However, recognizing
that the model we are simulating is rather simple, so should not
be expected to give highly precise quantitative results, our goal
in the present study is to be quantitatively only as accurate as
might be expected of such a simple model. Although we do not know
how accurate that really is, the general precision to which we
have chosen to work is roughly 10\% to 20\%. The quantitative
shifts due to our using a coarse grid in the simulations are at
most of this magnitude.  We have also verified that using a finer
grid spacing would not alter our results in any significant way.
Reduction of the grid spacing adds more degrees of freedom,
resulting in increased fluctuations, and suppression of $T_c$ and
$H_{c2}$. However, if one also lowers the value of the fluctuation
parameter
$\eta$ a little, this increase in the fluctuations can be removed,
and the resulting behavior of the system is only weakly dependent
on the grid spacing.

In the simulations, the time step is chosen to be between 0.02 and
0.1 in our units, with shorter steps at higher temperatures; we
check that our results are not affected by doubling the time step.
A first-order Euler method is used, with 10000 to 20000 time steps
used for equilibration, and 30000 to 50000 steps for averaging.
The two-dimensional data we show are for a 50$\times$50 square
grid. We use the gauge-invariance of the TDGL equation to work as
much as possible in terms of gauge-invariant quantities like
$|\Psi|$ and the gauge-invariant phase differences
\begin{equation}
\omega({\bf r,r'},t) = \theta({\bf r'},t) - \theta({\bf r},t)
-\int_{\bf r}^{\bf r'} {\bf A}\cdot{\bf dl} ~,
\end{equation}
where $\Psi=|\Psi|e^{i\theta}$.
The current densities are also written in terms of these quantities.

At $H=0$, for the two-dimensional system there is a
Kosterlitz-Thouless \cite{kosterlitz} transition at $T_c$, which
we locate by measuring the helicity modulus.

To obtain $\tilde{\alpha}_{xy}$, we measure the transverse energy current due an electric field.  We then use the Onsager relation to
obtain $\alpha_{xy}=\tilde{\alpha}_{xy}/T$.  We find that the signal to noise ratio for this measurement is significantly smaller than
in a direct measurement of $\alpha_{xy}$.  We have numerically verified the validity of this Onsager relation by measuring $\alpha_{xy}$
both ways.  This also serves as a check that we are indeed measuring the proper transport coefficients.  We also confirm that our
results agree with the analytic results \cite{iddo} in the higher-temperature regime where linearized TDGL applies.

The temperature enters the TDGL equation both in the term linear in the order parameter and in the intensity of the noise. It can be
shown that introducing a gradient in the linear term only adds to the magnetization currents, and thus for measuring the transport
currents it is sufficient to include the temperature gradient only in the noise term.  We have also verified this numerically.

\section{Results}

Some results for $\alpha_{xy}$ from the two-dimensional
simulations are shown in Fig. 1. We have chosen to compare these
results to the experimental results of Wang {\it et al.}
\cite{ong3} obtained from Nernst and resistivity measurements on
an overdoped LSCO sample with $x \cong 0.2$ and $T_c \cong 28$K.
For this comparison, we present the simulation results in SI units
using the parameters from the experiment, $H_{c2}(T=0) \cong 45$T
(thus $\xi \cong 27 \AA$) and layer spacing $s=6.6 \AA$.  The
fluctuation parameter has been adjusted to $\eta=0.42$ to give
reasonable quantitative agreement between simulation and
experiment. We set the temperature scale using $T_c^{MF}=40$K,
which gives $T_c \cong 28$K in the simulation. Note that
$\alpha_{xy}$ is not directly measured in the experiments.  For a
system with negligible Hall effect, $\alpha_{xy}$ can be obtained
from the measured Nernst effect and the longitudinal resistivity.
Thus for this comparison we require both these measurements to be
made on the same sample over a substantial portion of the vortex
liquid regime.  This requirement seriously limits the number of
experimental results we can compare to.

\begin{figure}[h!]
\begin{center}
$\begin{array}{c}
\epsfxsize=3in
\epsffile{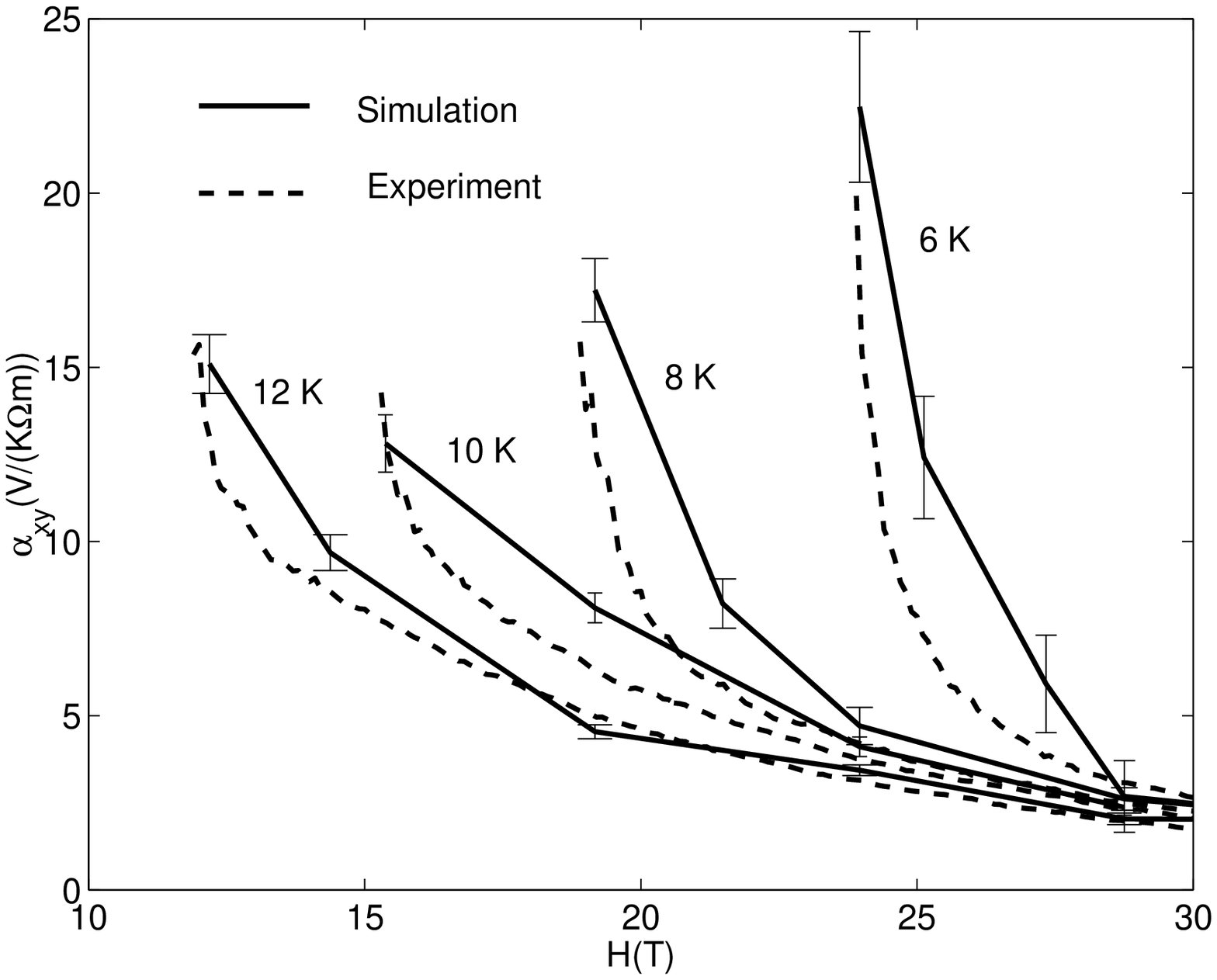}\\
\epsfxsize=3in
\epsffile{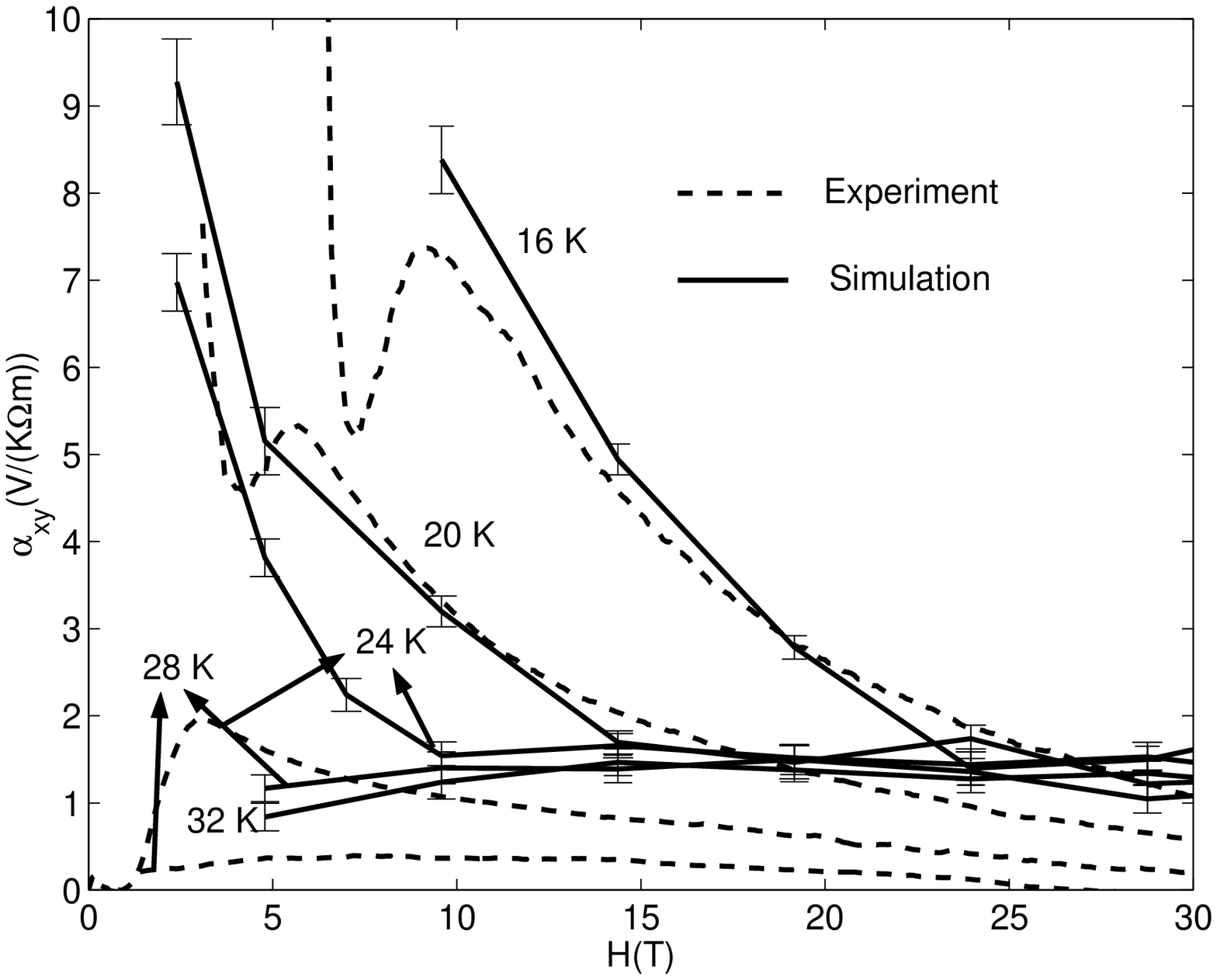} \\ 
\end{array}$
\end{center}
\caption{Experimental data obtained from measurements \cite{ong3}
of the Nernst effect on overdoped LSCO ($x=0.2$) along with the
results of our two-dimensional simulations.  The parameters used
in the simulation are: $T_{c}^{MF}=40$K, $H_{c2}(0)=45$T,
$s=6.6\AA$, $\eta=0.42$.  The critical temperature is $T_{c}
\approx 28$K in both simulation and experiment here. }
\label{experiment}
\end{figure}

Looking at Fig. 1, we see that there is reasonably good general
qualitative and quantitative correspondence between the simulation
and experiment, considering how simple the model is and that we
have adjusted only the strength of the thermal fluctuations $\eta$
to get this agreement. The other parameters are all dictated by
experiment.  Our results are thus consistent with the proposition
that the superconducting fluctuations modelled by the TDGL
equation produce most, if not all, of the contributions to the
large Nernst effect seen in the vortex liquid regime for this
cuprate sample.  Of course, the agreement between experiment and
this simple model is not perfect. One difference is seen at high
field in the higher-$T$ panel of Fig. 1, where the simulations
give a nearly-constant $\alpha_{xy}$, while the experiment shows
more variation.

One effect that is not included in our two-dimensional simulations
is interlayer coupling. For the cuprates, this can become
important in the vortex liquid at low fields in the vicinity of
$T_c$, where there is a crossover to three-dimensional behavior as
the correlation length normal to the layers becomes larger than
one layer spacing.  Note that the difference between simulation
and experiment is large at low fields and $T = 24$K. We considered
the possibility that this difference might be due to the crossover
to three-dimensional behavior, which is not present in our
two-dimensional simulation. To investigate this crossover, we have
also simulated the Lawrence-Doniach \cite{ld} version of the TDGL
equation.  In our units this is
\begin{widetext}
\begin{eqnarray}
(\partial_{t}+ i\phi)\Psi_{j} = (\bm{\nabla}_{\bot}-i{\bf A}_{\bot})^{2}\Psi_{j} - (T-1)\Psi_{j} 
  + J(e^{-isA_{z}}\Psi_{j+1} + e^{isA_{z}}\Psi_{j-1}-2\Psi_{j}) 
  + |\Psi_{j}|^{2}\Psi_j + \zeta ~, 
\end{eqnarray}
\end{widetext}
where $J$ is the interlayer Josephson coupling per unit area, and
$j$ is the layer index. The transition temperature is located by
finding the intersections of the fourth-order cumulant (Binder
ratio) curves
$\langle|\psi|^{4}\rangle/\langle|\psi|^{2}\rangle^{2}$ for
different system sizes as functions of temperature at zero
magnetic field.

\begin{figure}
\label{3d} \epsfxsize=3in \centerline{\epsfbox{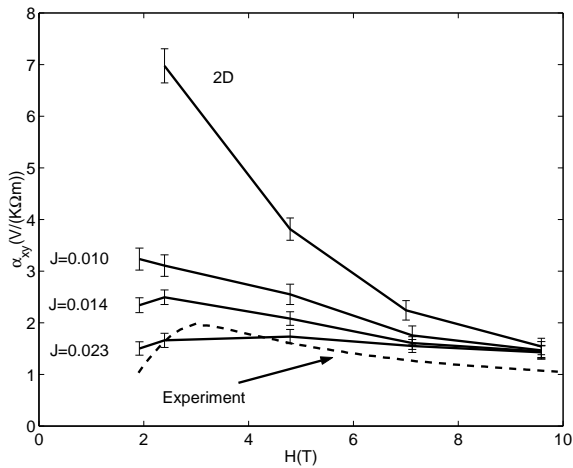}}
\caption{Three-dimensional simulation results for $T=24$K, with
interlayer coupling $J$.  Except for $T_c^{MF}$, the parameters
used here are the same as in Fig. 1.  See text for more details. }
\end{figure}

Fig. 2 shows results for $\alpha_{xy}$ as a function of field for
various values of interlayer coupling $J$ at temperature $T=24$K,
from samples of size 15$\times$15$\times$10 layers.  For each $J$
we set the the temperature scale ($T_c^{MF}$) so that $T_c \cong
28$K as in the experiment.
The values of the other parameters in the simulations are as used in Fig. 1.  As expected, the 
the largest value of $J$ shows the largest deviation from the
two-dimensional behavior.
Somewhat better correspondence between the behavior of
$\alpha_{xy}$ obtained in simulation and experiment is obtained by
adding this interlayer coupling.

An interesting point to note in Fig. 2 is that at low fields a rather small interlayer coupling $J$ produces a strong decrease in
$\alpha_{xy}$.  Describing this in terms of vortices, as seems appropriate in this low-field regime below $T_c$, the force per unit
length on a vortex line due to the temperature gradient is proportional to $\alpha_{xy}$.  Naively, this force is also set by the
entropy per unit length transported by a moving vortex.  This transport entropy may come from internal degrees of freedom within one
vortex, or it may come from the configurational entropy of the many possible spatial arrangements of the vortices.  The latter does not
involve internal excitations within the vortices.  In the quasi-two-dimensional layered superconductors that we are modelling, a vortex
line running normal to the layers consists of ``pancake vortices" in each layer.  A weak interlayer coupling has little effect on the
internal properties of a single pancake vortex, but it does produce an attraction between vortices in adjacent layers that reduces their
relative motion, and thus can substantially reduce the configurational entropy of the vortices.  The fact that at low fields a weak
interlayer coupling greatly reduces the transport entropy of the vortices, as indicated by the reduction of $\alpha_{xy}$ seen in Fig.2,
thus suggests that in this regime the configurational entropy of the vortices is a large part of their transport entropy, at least
within this layered TDGL model. 

We have also performed simulations to model the BSCCO sample
studied by Ri {\it et al.} \cite{ri}. The experimental data and
the results of the simulation are shown in Fig. 3. The quantity
plotted is the measured ``transport energy'' of the vortices,
defined as $U_{\phi} = T\phi_{0}\alpha_{xy}$.  We used a
two-dimensional simulation with $T_{c} \cong 90$ K, $H_{c2} = 160$
T, $s=15.35 \AA$ and $\eta = 0.475$. Again, there is reasonable
quantitative agreement, in absolute units, between the
experimental data and the simulations. The clearest deviation here
occurs for $T<60$K and $H>6$T,
\begin{widetext}
\begin{figure}
$\begin{array}{cc}
\epsfxsize=2.5in
\epsfysize=3in
\epsffile{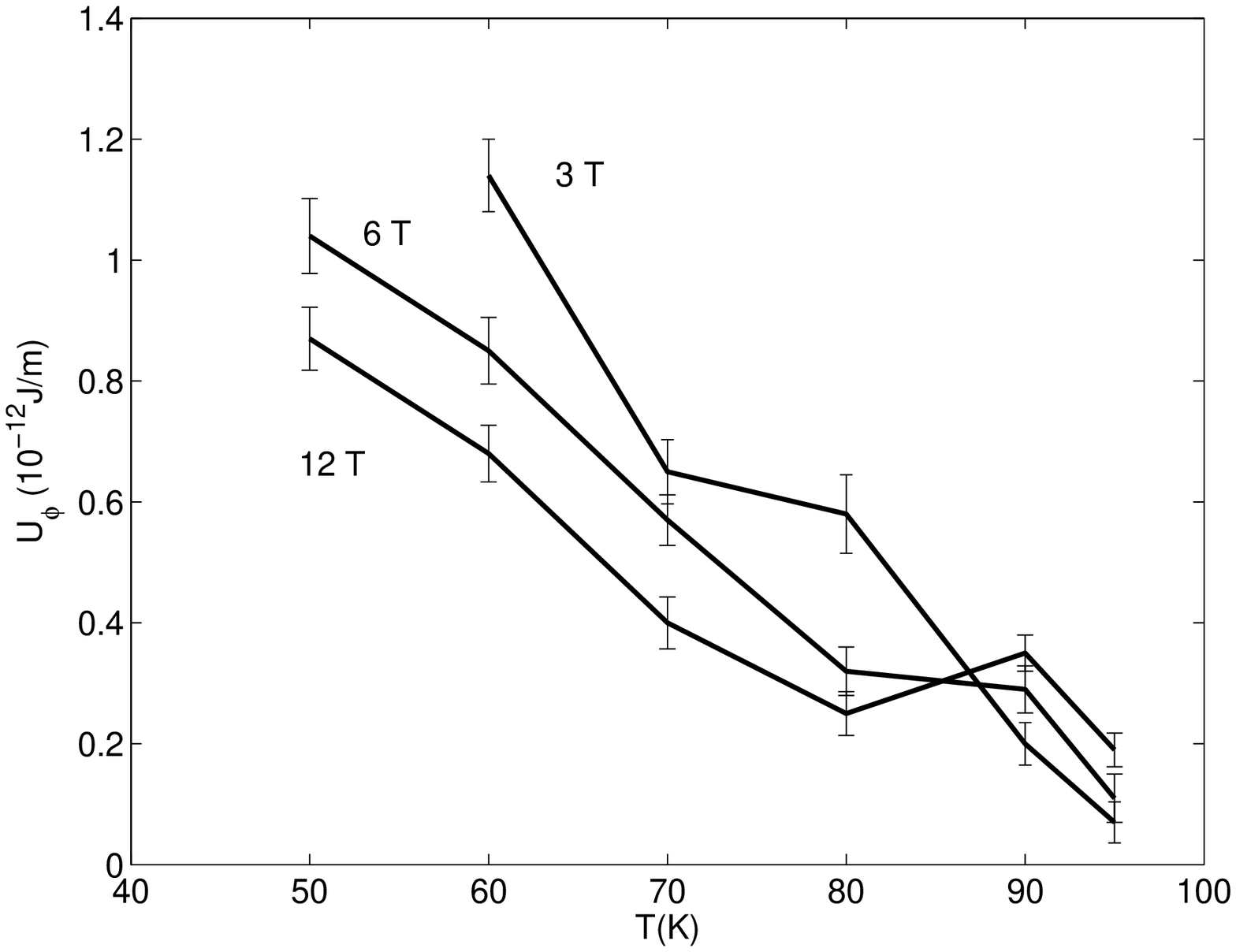} &
\epsfxsize=3in
\epsffile{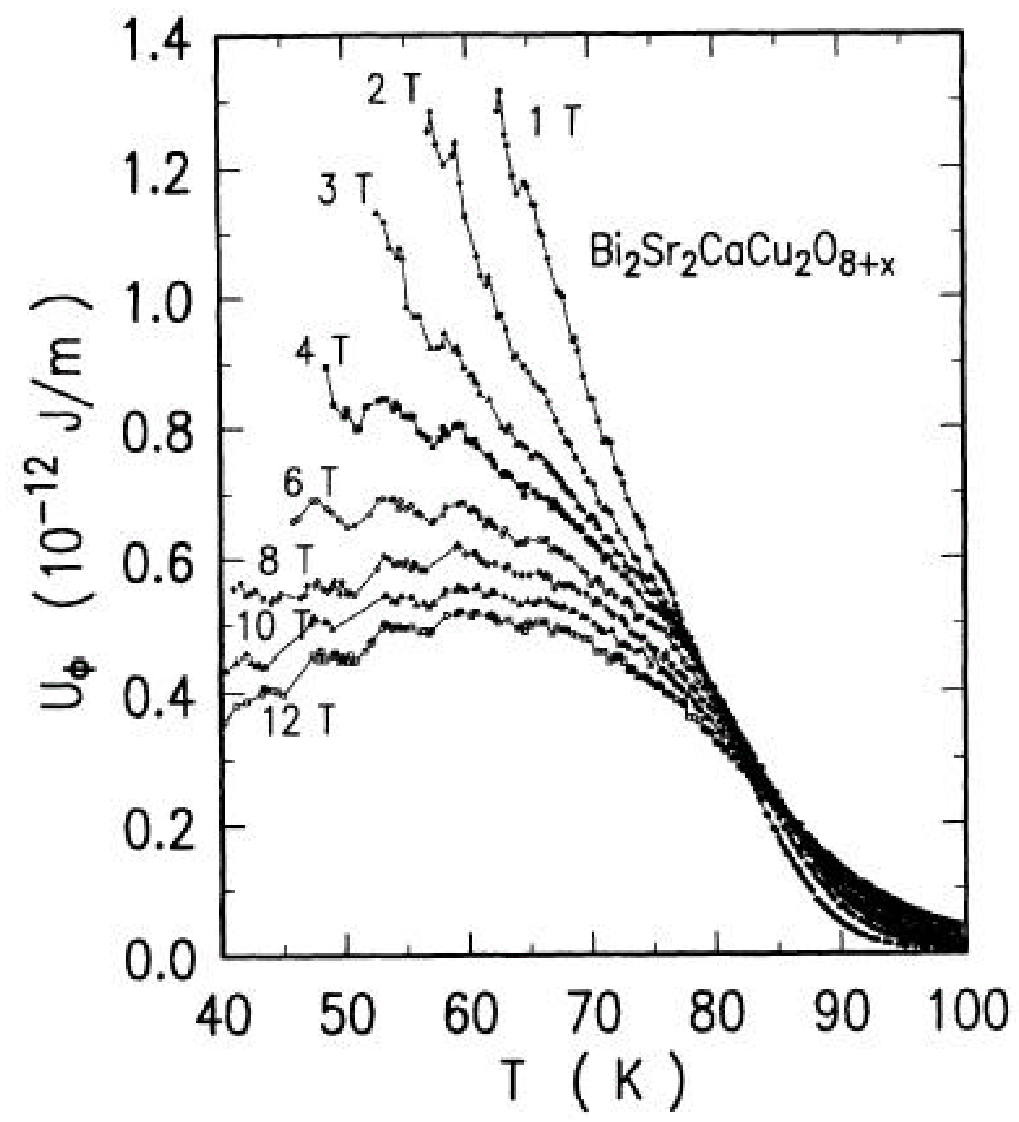} \\[0.3cm]
\end{array}$
\caption{(Left) Data from the simulation with $s = 15.35 \AA$,
$T_{c} = 90$ K, $H_{c2} = 160$ T and $\eta=0.475$.
(Right) Experimentally measured \cite{ri} value of
$U_{\phi}$ obtained from Nernst measurements on optimally doped
BSSCO (2212) with $T_{c}$= 90 K and $H_{c2}(0) \approx$ 160 T.}
\end{figure}
\end{widetext}
where the sign of the temperature derivative of $U_{\phi}$ appears
to differ between simulation and experiment.


In conclusion, we have
measured the transverse thermoelectric transport coefficient
$\alpha_{xy}$ in simulations of the vortex-liquid regime of
superconductors modeled by the TDGL equation with thermal
fluctuations.
We find that our simulations of a two-dimensional superconductor
reproduce reasonably well much of the qualitative and quantitative
features of available experimental data from some cuprate
high-temperature superconductors.  We have also studied the
crossover from two-dimensional to three-dimensional behavior in a
layered superconductor.  This crossover study indicates that the
configurational entropy of the vortices may constitute a large
part of their transport entropy at low magnetic field just below
$T_c$.

\begin{acknowledgments}
We thank Yayu Wang and N. P. Ong for providing us with the LSCO data.  We also thank them, Shivaji Sondhi and especially Iddo Ussishkin
for many useful discussions.  This work is supported by the NSF through MRSEC grant DMR-0213706.
\end{acknowledgments}


\end{document}